\documentclass[12pt]{article}

\usepackage{newtxtext,newtxmath}
\usepackage[T1]{fontenc}
\usepackage{amsmath}
\usepackage{graphicx}
\usepackage{array}
\usepackage[hidelinks]{hyperref}
\usepackage{bm}
\usepackage{setspace}
\usepackage[top=1in, bottom=1in, right=1in, left=1in]{geometry}
\usepackage[authordate, autocite=inline, backend=bibtex, natbib]{biblatex-chicago}
\usepackage{lettrine}
\usepackage{enumitem}
\usepackage{authblk}
\usepackage{sectsty}
\allsectionsfont{\sffamily}
\usepackage{subcaption}
\DeclareCaptionFormat{custom}{\textsf{\textbf{#1#2}#3}}
\usepackage{abstract}

\usepackage{adjustbox}
\usepackage{caption}

\title{\textsf{\textbf{Uncovering the Effect of Toxicity on Player Engagement and its Propagation in Competitive Online Video Games}}}
\author{Jacob Morrier\thanks{Corresponding author. E-mail address: \href{mailto:jmorrier@caltech.edu}{jmorrier@caltech.edu}.}}
\affil{Division of the Humanities and Social Sciences, California Institute of Technology}
\author{Amine Mahmassani}
\affil{Activision\textregistered}
\author[1]{R. Michael Alvarez}
\date{July 2024}

\begin{document}

\maketitle

\begin{abstract}
    This article seeks to provide accurate estimates of the causal effect of exposure to toxic language on player engagement and the proliferation of toxic language. To this end, we analyze proprietary data from the first-person action video game \emph{Call of Duty\textregistered: Modern Warfare\textregistered III}, published by Activision\textregistered. To overcome causal identification problems, we implement an instrumental variables estimation strategy. Our findings confirm that exposure to toxic language significantly affects player engagement and the probability that players use similar language. Accordingly, video game publishers have a vested interest in addressing toxic language. Further, we demonstrate that this effect varies significantly depending on whether toxic language originates from opponents or teammates, whether it originates from teammates in the same party or a different party, and the match’s outcome. This has meaningful implications regarding how resources for addressing toxicity should be allocated.
\end{abstract}

\bigskip

\noindent \textsf{\textbf{Keywords:}} causal inference, competitive online video games, instrumental variables estimation, propagation, player engagement, toxicity.

\clearpage

\section{Introduction}

This article seeks to provide accurate estimates of the causal effect of exposure to toxicity on player engagement and the probability that players engage in similar behavior across various contexts. These estimates are of great interest to both social scientists and video game publishers for at least three reasons.

First, these estimates can help formulate a business case for addressing toxicity. Video game publishers may address toxicity for several reasons. These may include protecting players from psychological harm and promoting an inclusive and positive gaming environment. Moreover, if exposure to toxicity is detrimental to player engagement, addressing it can mitigate its harmful effects. This is particularly compelling because, in such cases, video game publishers have a vested interest in addressing toxicity, as it can directly affect their products’ performance.

Once the business case for addressing toxicity has been embraced, the effect of exposure to toxicity and, particularly, its variations across different contexts carry meaningful implications regarding the distribution of resources for combating it. With limited resources available to address toxicity, one should allocate them where they will have the greatest impact. Therefore, priority should be given to addressing toxicity in the contexts where its undesirable effects are most pronounced. This ensures that every prevented instance of toxicity yields the highest returns in terms of improved player engagement. In contrast, efforts should be diverted from contexts where players may derive gratification in behavior otherwise considered toxic. By investigating the impact of such behavior on player engagement, publishers can better distinguish toxic behavior from acceptable conduct and focus their resources on addressing the former.

Finally, understanding how toxicity propagates and causes other players to engage in similar behavior is instrumental in advancing the goal of reducing exposure to toxicity, regardless of its effect on player engagement. An often-repeated aphorism is that humans are, by nature, social animals. Accordingly, their behavior tends to be strongly influenced by their peers. Countless empirical studies, experimental and observational, document strong correlations and causal relationships between an individual's behavior and outcomes and those of their peers \citep{Manski_2000, ALEXANDER200122, SALMIVALLI2010112, EPPLE20111053, Kreager_et_al_2011, SACERDOTE2011249, Graham_2018}. This is equally true of virtuous and reprehensible behavior, including academic cheating, bullying, and delinquency. Therefore, we expect toxicity to proliferate, meaning that one player engaging in toxicity causes the exposed players to engage in similar behavior. In contexts where this proliferation is more pronounced, allocating additional resources to address toxicity will generate greater returns mechanically.

To achieve our ambition, we analyze proprietary data from the first-person action video game franchise \emph{Call of Duty\textregistered}, published since 2003 by the leading global interactive entertainment company Activision\textregistered. We focus on the franchise’s latest installment, \emph{Call of Duty: Modern Warfare\textregistered III}. In its primary multiplayer game mode, Team Deathmatch, players are divided into two equally sized teams. Players' objective is to eliminate players from the opposing team as often as possible. When a player is eliminated, they reappear at a different location on the map. The team that first hits a predetermined point limit or accumulates the most eliminations by the end of the game wins. A draw is declared if both teams have the same score at the end of the game.

Since the summer of 2023, Activision has partnered with Modulate\texttrademark, a start-up company developing intelligent voice technology to combat online toxicity and foster healthy and safe online communities \citep{toxmod_description}. Activision has incorporated Modulate's proprietary voice chat moderation technology, ToxMod\texttrademark, into its gaming platforms. ToxMod uses artificial intelligence and machine learning algorithms to detect toxic language in real-time, including discriminatory language, harassment, and hate speech.\footnote{For more details on ToxMod’s functionalities, see \citet{kowert_woodwell}.} The initial beta rollout of this voice chat moderation technology began in North America on August 30, 2023, within the games \emph{Call of Duty: Modern Warfare II} and \emph{Call of Duty: Warzone\texttrademark}. This was followed by a global release coinciding with the launch of \emph{Call of Duty: Modern Warfare III} on November 10, 2023. Support was initially provided in English, with plans to extend to additional languages. We use the data produced by ToxMod to conduct our analysis. Our dataset contains all observations from the matches monitored by ToxMod from November 10 to December 10, 2023, corresponding to the first month immediately following the game's release. We classify a player as having used toxic language in a match if ToxMod identified at least one of their statements during it as toxic.

We conduct two sets of regression analyses. The first considers the effect of exposure to toxic language from opponents and teammates conditional on whether the player’s team wins or loses the match. The second considers the effect of exposure to toxic language from teammates in a different party and the same party---teammates with whom they were algorithmically assigned or with whom they intentionally teamed up, respectively---conditional on whether the player’s team wins or loses the match.\footnote{Parties are groups of one or more players who have chosen to play together. The matchmaking algorithm usually keeps parties together when forming teams.} In both cases, we consider the effect of exposure to toxic language on two outcomes: (i) the time it takes a player to enter their next match, reflecting short-term player engagement, and (ii) the probability that they use to toxic language in the current game, reflecting the immediate propagation of toxic language.

Even with a large amount of high-quality data, analysts seeking to estimate the causal effect of exposure to toxic language on player engagement and the probability of using such language face considerable statistical challenges. The reason is that some variables not controlled for in our regression models---because they are unmeasured, for instance---may simultaneously impact players’ outcomes and whether they are exposed to toxic language. For example, teammates might concomitantly use toxic language in reaction to a random event occurring in a match. This random event may simultaneously influence players’ short-term player engagement. More fundamentally, players mutually affect each other. Consequently, whether a player, their teammates, and their opponents engage in toxicity are jointly determined. This obscures the cause-to-effect relationship of exposure to toxic language and introduces biases in standard ordinary least squares (OLS) estimates.

To overcome this causal identification problem, we put forth an identification strategy leveraging the fact that we observe players partaking in multiple matches with different players. Through an instrumental variable or two-stage least squares (2SLS) estimation strategy, we isolate variations in outcomes of interest caused by interactions with players who, in prior matches with other players, have employed toxic language more frequently and, consequently, are more likely to use toxic language in the current game. This approach allows us to reliably assess whether and, if so, to what extent exposure to toxic language \emph{causes} variations in player engagement and the probability of using similar language.

\section{Methods}

\subsection{Dataset Description}

Our dataset contains data from all the matches in Team Deathmatch mode monitored by ToxMod in \emph{Call of Duty: Modern Warfare III} from November 10 to December 10, 2023. This amounts to 56,464,489 observations, each representing a player in a match, from 4,167,325 matches and 4,539,599 players.\footnote{Our sample is not comprehensive, as it only includes matches monitored through ToxMod in a single game mode.} On average, we observe each player participating in 12.44 games.

\begin{figure}[!p]
    \centering
    \includegraphics[width=0.9\linewidth]{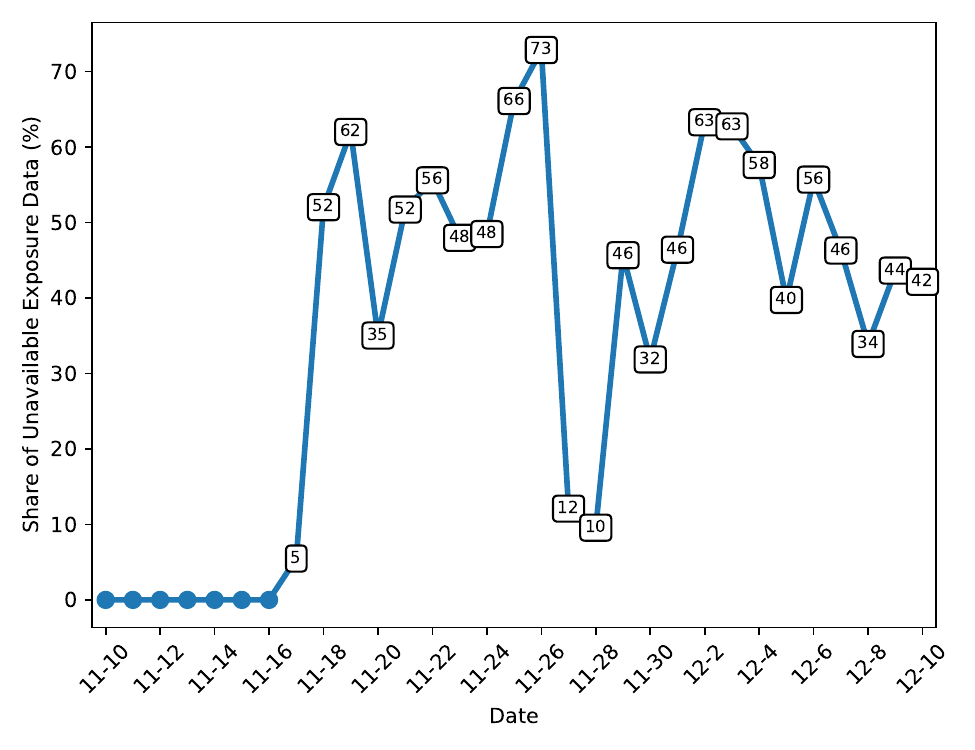}
    \caption{Daily Evolution of the Share of Unavailable Exposure Data}
    \label{fig:missing_exposure_data}
\end{figure}

Because of technical reasons, exposure data is unavailable for some ToxMod offenses. Data on which players were exposed is unavailable for 34.6\% of the statements identified as containing toxic language by ToxMod in the matches considered in our analysis. Figure \ref{fig:exposure} illustrates the daily evolution of the share of unavailable exposure data over our period of interest. From November~17, one week after the game's launch, exposure data for some offenses is unavailable. Over the remainder of the period of study, the daily share of unavailable exposure data fluctuated arbitrarily between five and 73\%, with 35 to 65\% of exposure data unavailable on most days.

\begin{figure}[!p]
    \centering
    \begin{subfigure}[b]{0.75\linewidth}
        \centering
        \includegraphics[width=\linewidth]{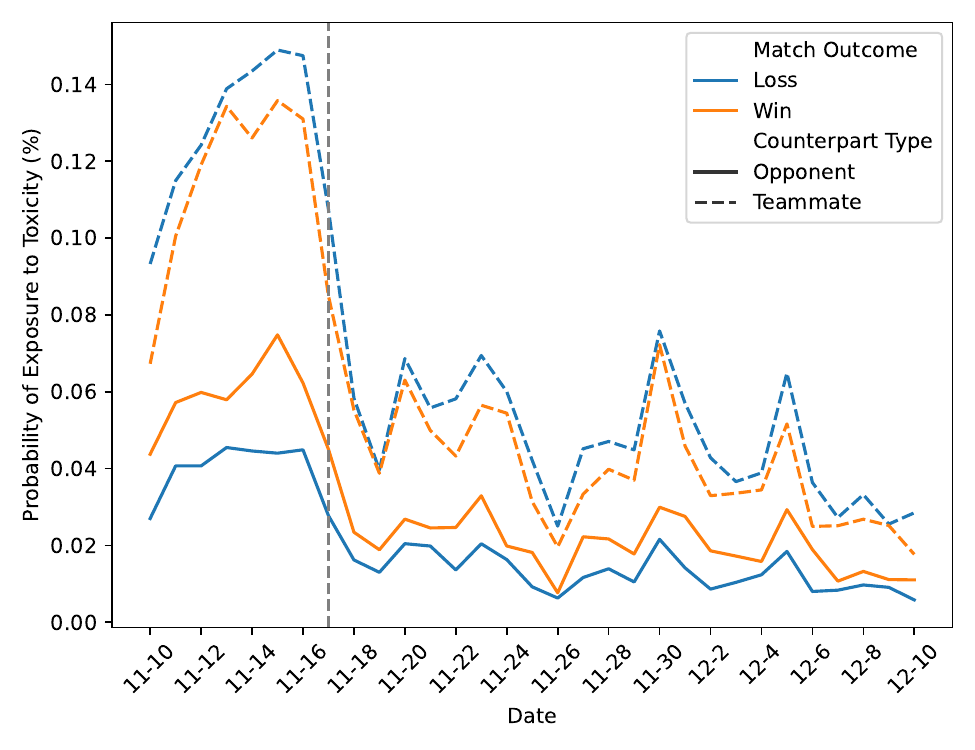}
        \caption{Opponents and Teammates}
    \end{subfigure}

    \begin{subfigure}[b]{0.75\linewidth}
        \centering
        \includegraphics[width=\linewidth]{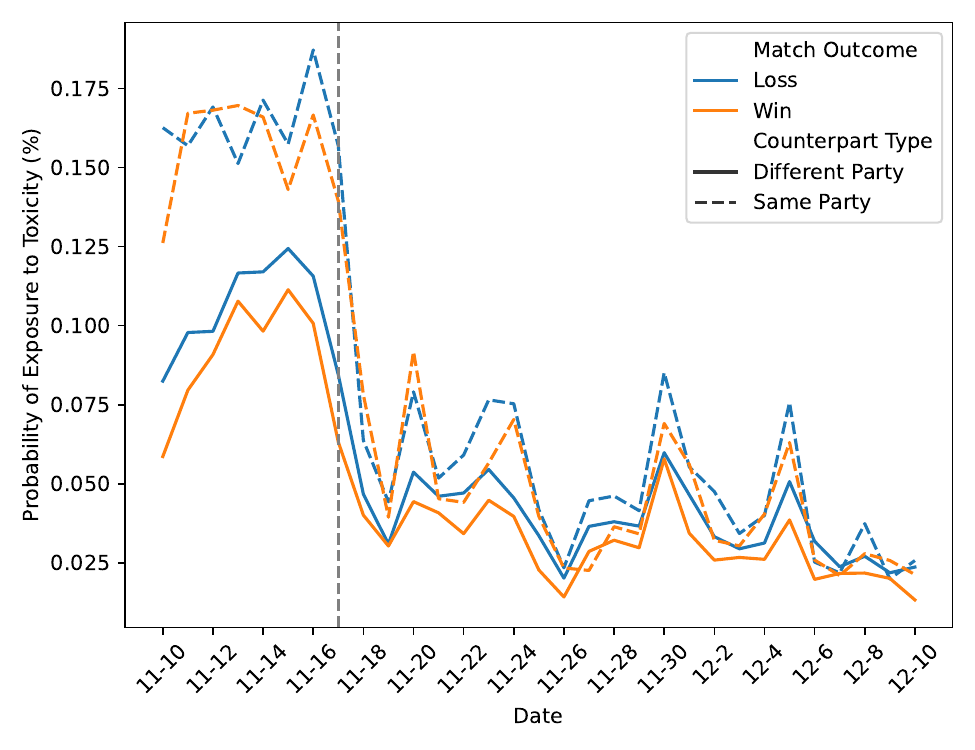}
        \caption{Teammates in a Different Party and the Same Party}
    \end{subfigure}
    \caption{Daily Evolution of the Probability of Exposure to Toxic Language}
    \label{fig:exposure_time}
\end{figure}

\begin{figure}[!p]
    \centering
    \begin{subfigure}[b]{0.9\linewidth}
        \centering
        \includegraphics[width=\linewidth]{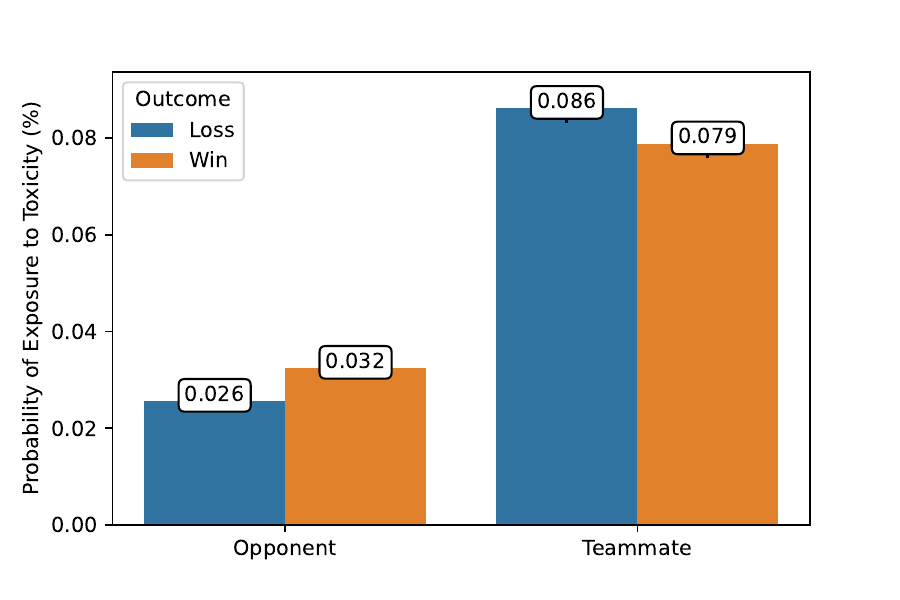}
        \caption{Opponents and Teammates}
    \end{subfigure}

    \begin{subfigure}[b]{0.9\linewidth}
        \centering
        \includegraphics[width=\linewidth]{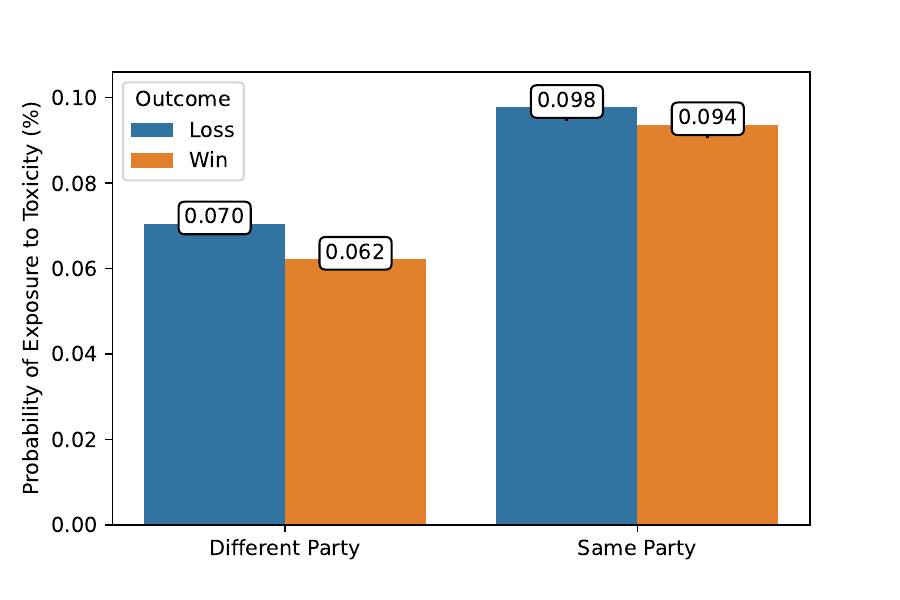}
        \caption{Teammates in a Different Party and the Same Party}
    \end{subfigure}
    \caption{Probability of Exposure to Toxic Language from March 4 to April 12, 2024}
    \label{fig:exposure_feb_apr}
\end{figure}

If exposure data is randomly unavailable and, specifically, unavailable independently of the outcomes of interest, it will not introduce any bias in our findings.\footnote{In fact, it may dilute the magnitude of our coefficients. However, since the probability of exposure to toxic language is small, this dilution is most likely negligible.} Although proving that exposure data is randomly unavailable is challenging, it appears probable given the known cause of the data availability disruption. To make the case for this conjecture, we present two pieces of evidence showing that exposure data is unavailable in roughly similar proportions depending on whether the toxic language emanated from an opponent or a teammate, whether it emanated from a teammate in a different party or the same party and whether the player’s team won or lost.

First, Figure \ref{fig:exposure_time} illustrates the daily evolution of the probability that players are exposed to toxic language depending on the context over our period of study. The day after which some exposure data is unavailable is marked using a dashed vertical line. In general, the probability that a player is exposed to toxic language in different contexts varies proportionally over this period. Accordingly, the relationships between exposure probabilities across contexts are roughly preserved throughout our period of study. More importantly, the proportions between exposure probabilities across contexts are preserved between the period for which we have comprehensive exposure data---before November 17---and the period for which exposure data is missing.

Second, Figure \ref{fig:exposure_feb_apr} illustrates the probability that a player is exposed to toxic language from opponents or teammates in a different party or the same party based on whether the player’s team won or lost for the period from March 4 to April 12, 2024, for which we have comprehensive exposure data for a randomly drawn subset of matches. This figure shows that the observed exposure to toxic language across various contexts, as illustrated in Figure \ref{fig:exposure}, occurs in approximately the same proportions over our period of study compared to a subsequent period for which we have comprehensive exposure data, although slightly smaller in magnitude.

\subsection{Model Specification}

In this article, we seek to estimate the causal effect exposure to toxic language exerts on player engagement and the probability of using similar language. We define these effects as the variation in the average time it takes a player to enter their next match and the probability that they use toxic language in the current game, respectively, caused by exposure to toxic language from another player, holding all other variables constant.

To this end, we formulate the following structural model of players' behavior:
\[
    y_{ij} = \alpha_{j} + \bm{\beta} \cdot \bm{x}_{ij} + \varepsilon_{ij}.
\]

In this model, $y_{ij}$ represents the outcome of interest in match $i$ for player $j$, that is, either the average time it takes a player to enter their next match or the probability that they use toxic language in the current game, $\alpha_{j}$ a player-specific intercept, $\bm{x}_{ij}$ a vector including the number of players other than player $j$ who have used toxic language in match $i$ and to which player $j$ was exposed, and $\varepsilon_{ij}$ an error term. The vector $\bm{x}_{ij}$ also contains interactions between the number of other players using toxic language in a match and whether player $j$’s team has won or lost the game. We define a player as having used toxic language in a match if ToxMod recognized at least one of their statements as toxic.

In essence, our structural model of players' behavior postulates that the outcomes of interest are primarily affected by two factors: (i) their inherent tendency to display this outcome, and (ii)~the number of other players, either teammates or opponents or teammates in a different party or the same party, who used toxic language to which they were exposed. The coefficients $\bm{\beta}$ capture the effect of exposure to toxic language on outcomes of interest. They are the estimand of our analysis.

Before delving into causal identification issues, let us discuss what the time it takes a player to enter their next match reflects. Consider a player who ends their current session and plans to return at the same hour the next day. In this case, 24 hours will elapse until this player’s next match. Conversely, if a player continues playing and immediately joins a new match, there will be a near-zero time until their next match. In essence, the time it takes a player to enter their next match reflects the interaction of two things: (i) the likelihood that they will end their current playing session at the end of the match, and (ii) the time it takes for them to return and initiate a new session.

\subsection{Causal Identification Problems} 

Naturally, one might contemplate estimating the coefficients $\bm{\beta}$ through OLS. However, contrary to standard assumptions in linear regression models, the explanatory variables are not independent of the error terms. This ultimately results in endogeneity.

There can be several causes of endogeneity, each posing a unique threat to the causal identification of our estimands. For instance, endogeneity may arise because the model is misspecified and, more precisely, some variables not included in the structural model of players’ behavior---because they are not measured, for instance---simultaneously affect the outcomes of interest and whether the player is exposed to toxicity. Concretely, endogeneity may arise because a player and their teammates might both resort to toxic language in reaction to an exogenous event happening in the game. This random event may also affect how long it takes for players to enter their next match.

Another threat to causal identification is self-selection. It is common for players to form ``parties,'' enabling them to team up in matches. Players may form such parties to intentionally engage in toxicity or in the anticipation that their fellow party members will engage in such behavior. Also, when two players decide to team up, it suggests that they already know each other well. This familiarity is likely to change the dynamics of their relationship, including the probability of using toxic language and being exposed to such language through one another. It may concurrently affect player engagement, causing them to enter their next match more quickly when paired. Even if players do not deliberately join forces, their previous interactions are likely to have significant enduring effects.

Endogeneity also emerges mechanically when we seek to estimate the effect of exposure to toxic language on the probability that a player uses toxic language of their own. This stems from the fact that players in a match mutually affect each other. To see this, consider the simplistic case of a player with a single teammate and no opponent. In this scenario, the dependent variable in some equations appears on the right-hand side of others. This reflects that the player and their teammate mutually influence each other, and whether they employ toxic language is jointly determined. It follows that the variable indicating whether a player's teammate engages in toxic language is endogenous.

To see this formally, let us consider the pair formed by players $j$ and $k$ in match $i$. The two equations governing whether these players engage in toxic speech are as follows:
\begin{equation*}
    \begin{split}
        Y_{ij} & = \alpha_{j} + \beta \times Y_{ik} + \varepsilon_{ij} \\
        Y_{ik} & = \alpha_{k} + \beta \times Y_{ij} + \varepsilon_{ik}.
    \end{split}
\end{equation*}
    
To prove that $Y_{ik}$ is correlated with $\varepsilon_{ij}$, it suffices to incorporate the first equation into the second one and rearrange the output to isolate $Y_{ik}$ on the left-hand side:
\begin{equation*}
    \begin{split}
        Y_{ik} = \alpha_{k} + \beta \times \left(\alpha_{j} + \beta \times Y_{ik} + \varepsilon_{ij} \right) + \varepsilon_{ik} &  \Leftrightarrow \left(1 - \beta^{2}\right) \times Y_{ik} = \alpha_{k} + \beta \times \left(\alpha_{j} + \varepsilon_{ij}\right) + \varepsilon_{ik} \\
        & \Leftrightarrow Y_{ik} = \frac{\beta}{1 - \beta^{2}} \times \left(\alpha_{j} + \varepsilon_{ij}\right) + \frac{1}{1 - \beta^{2}} \times \left(\alpha_{k} + \varepsilon_{ik}\right).
    \end{split}
\end{equation*}
    
This equation implies that the error term $\varepsilon_{ij}$ directly enters the value of $Y_{ik}$, resulting in a correlation between them. Intuitively, this means that OLS estimates do not only capture a teammate’s effect on a player’s inclination to engage in toxic speech but also its ``reflection,'' that is, the effect that player has on their teammate.

\subsection{Identification Strategy}

To overcome the causal identification problem described above, we outline an identification strategy that leverages the fact that we observe players participating in multiple matches with different players. Our proposed approach is to implement an instrumental variable or 2SLS estimation strategy. We propose to instrument the variables representing the number of teammates and opponents using toxic language to which the player was exposed in the current match with the sum of their individual probabilities to have used such language \emph{in prior matches with other players}. This identification strategy isolates variations in outcomes of interest caused by interactions with players who, in previous matches with other players, have had a greater tendency to use toxic language and, consequently, are more likely to use such language in the current game.

Hereafter, for simplicity, we consider a model in which exposure to toxic language is treated the same regardless of whether it originates from teammates or opponents. In this model, there is a single coefficient capturing the average effect of one other player using toxic language to which the player is exposed. Our approach can be straightforwardly extended to the case in which we consider exposure to toxicity from opponents and teammates and from teammates in a different party and the same party separately.

Formally, our identification strategy consists of adding the following equation to our model of players’ behavior:
\[
    x_{ij} = \delta_{j} + \gamma \times \sum_{k \in \mathcal{P}_{i, -j}} \sum_{\ell \in \mathcal{M}_{i, k, -j}} \frac{x^{\star}_{\ell k}}{\# \mathcal{M}_{i, k, -j}}+ u_{ij},
\]
where $\mathcal{P}_{i, -j}$ is the set of players in match $i$ excluding player $j$, $\mathcal{M}_{i, k, -j}$ the set of matches prior to match $i$ to which player $k$ participated but not player $j$, and $x^{\star}_{\ell k}$ a variable indicating whether player $k$ used toxic language in match $\ell$. The instrument is computed by summing over all other players in match $i$, indexed by $k$, the probability with which they have used toxic language in the previous matches they participated in but player $j$ did not, indexed by $\ell$. This instrumental variable belongs to the general class of spatial or ``leave-one-out'' instruments pioneered in empirical industrial organization for demand and supply estimation \citep{Hausman_1996, nevo_2001}.

\begin{table}[!p]
    \centering
    \caption{First-Stage Regression Results}
    \begin{subtable}[b]{\linewidth}
        \centering
        \caption{Opponents and Teammates}
        \footnotesize
        \begin{tabular}{lcccc}
            \hline
            \hline
            \\[-1.5ex]
             & \textbf{Opponents} & \textbf{Teammates} & \textbf{Opponents $\times$ Win} & \textbf{Teammates $\times$ Win} \\
            \\[-1.5ex]
             \hline
            \\[-1.5ex]
             Opponents & 0.0082$^{***}$ & 0.0015$^{***}$ & $-$0.0003$^{***}$ & $-$0.0005$^{***}$ \\
             & (0.000) & (0.000) & (0.000) & (0.000) \\
            \\[-1.5ex]
             Teammates & $-$0.0001 & 0.0330$^{***}$ & $-$0.0003$^{***}$ & $-$0.0012$^{***}$ \\
             & (0.000) & (0.001) & (0.000) & (0.000) \\
            \\[-1.5ex]
             Opponents $\times$ Win & 0.0047$^{***}$ & $-$0.0012$^{***}$ & 0.0156$^{***}$ & 0.0014$^{***}$ \\
             & (0.001) & (0.000) & (0.001) & (0.000) \\
            \\[-1.5ex]
             Teammates $\times$ Win & 0.0007$^{**}$ & $-$0.0012 & 0.0009$^{***}$ & 0.0345$^{***}$ \\
             & (0.000) & (0.001) & (0.000) & (0.001) \\
            \\[-1.5ex]
             Win & 0.0001$^{***}$ & 0.0000$^{***}$ & 0.0002$^{***}$ & 0.0005$^{***}$ \\
             & (0.000) & (0.000) & (0.000) & (0.000) \\
            \\[-1.5ex]
            \hline
            \\[-1.5ex]
            $F$ Statistic & 365.2 & 801.8 & 1615.0 & 3159.9 \\
            \\[-1.5ex]
            \hline
            \hline
            \\[-2ex]
            \textit{Note:}  & \multicolumn{4}{r}{$^{*}$p$<$0.1; $^{**}$p$<$0.05; $^{***}$p$<$0.01} \\ 
        \end{tabular}
    \end{subtable}

    \bigskip

    \begin{subtable}[b]{\linewidth}
        \centering
        \caption{Teammates in a Different Party and the Same Party}
        \footnotesize
        \begin{adjustbox}{center}
        \begin{tabular}{lcccc}
            \hline
            \hline
            \\[-1.5ex]
             & \textbf{Different Party} & \textbf{Same Party} & \textbf{Different Party $\times$ Win} & \textbf{Same Party $\times$ Win} \\
            \\[-1.5ex]
             \hline
            \\[-1.5ex]
             Different Party & 0.0271$^{***}$ & $-$0.0001$^{*}$ & $-$0.0003$^{***}$ & $-$0.0001$^{***}$ \\
             & (0.001) & (0.000) & (0.000) & (0.000) \\
            \\[-1.5ex]
             Same Party & $-$0.0010$^{**}$ & 0.0359$^{***}$ & $-$0.0004$^{**}$ & $-$0.0045$^{***}$ \\
             & (0.000) & (0.003) & (0.000) & (0.001) \\
            \\[-1.5ex]
             Different Party $\times$ Win & $-$0.0002 & $-$0.0001 & 0.0275$^{***}$ & $-$0.0000 \\
             & (0.001) & (0.000) & (0.001) & (0.000) \\
            \\[-1.5ex]
             Same Party $\times$ Win & $-$0.0002 & $-$0.0005 & $-$0.0002 & 0.0449$^{***}$ \\
             & (0.001) & (0.004) & (0.000) & (0.003) \\
            \\[-1.5ex]
             Win & $-$0.0001$^{***}$ & $-$0.0000$^{***}$ & 0.0004$^{***}$ & 0.0001$^{***}$ \\
             & (0.000) & (0.000) & (0.000) & (0.000) \\
            \\[-1.5ex]
             Player Belong to Party & $-$0.0000$^{*}$ & 0.0007$^{***}$ & $-$0.0000 & 0.0003$^{***}$ \\
             & (0.000) & (0.000) & (0.000) & (0.000) \\
            \\[-1.5ex]
            \hline
            \\[-1.5ex]
            $F$ Statistic & 505.1 & 917.4 & 2051.4 & 605.2 \\
            \\[-1.5ex]
            \hline
            \hline
            \\[-2ex]
            \textit{Note:}  & \multicolumn{4}{r}{$^{*}$p$<$0.1; $^{**}$p$<$0.05; $^{***}$p$<$0.01} \\ 
        \end{tabular}
        \end{adjustbox}
    \end{subtable}
    \label{tab:first_stage}
\end{table}

In general, for an instrumental variable to be valid, it must meet two conditions: (i) relevance, meaning that the instrumental variables must be strongly correlated with the endogenous explanatory variables, and (ii) exclusion, meaning that the instrumental variables must be independent of the error term of the structural equation. The validity of the first condition can be verified empirically by examining the first-stage regression. As a rule of thumb, the $F$ statistic against the null hypothesis that the instruments are irrelevant in the first-stage regressions should have a value greater than ten. Table \ref{tab:first_stage} presents the coefficients associated with the instrumental variables and the exogenous explanatory variables and the $F$ statistic for all first-stage regressions in our analysis. Each column represents an endogenous explanatory variable, and each row an instrumental variable or exogenous explanatory variable. In each case, the $F$ statistic is substantially greater than ten, suggesting that we undoubtedly have a ``strong first stage.''

On the other hand, the validity of the exclusion restriction cannot be empirically tested. Instead, it hinges on the assumptions we are willing to make regarding the relationship between the instrumental variables and the structural equation’s error term. We argue that using other players’ individual probabilities of having used toxic language in previous matches with other players to compute the instrument neutralizes the principal sources of endogeneity and, consequently, ensures that the exclusion restriction holds.

First, the fact that no data from the current match enters the computation of the instrumental variables neutralizes endogeneity caused by events occurring in the current match that simultaneously affect the outcomes of interest and exposure to toxic language. For instance, it resolves the case in which a player and one or more of their teammates use toxic language in reaction to, say, one of their common opponents using such language.

Second, the fact that no data from the other matches in which both players participated enters the computation of the instruments neutralizes endogeneity from enduring factors reflecting their relationship and simultaneously affecting outcomes of interest and whether they are exposed to toxic language, particularly through each other.

Third, using only data from past matches in the computation of the instrumental variables neutralizes the long-term effects of exposure to toxic language and other variables on the outcomes of interest. This is especially vital when estimating the effect of exposure to toxic language on a player’s probability of using similar language. Indeed, whether player $j$ uses toxic language in a match may affect the propensity of one of the other players, say, player $k$, to employ such language in future matches, irrespective of whether player $j$ participates in it or not. More generally, the events occurring in the current game may have a lasting impact on players’ behavior in the future. If data from future matches entered the computation of the instrumental variables, it would open a backdoor for a player’s use of toxic language or other events in the current game to affect the instrument, leading to a direct infringement of the exclusion restriction.

In interpreting our findings, it is essential to keep in mind that our estimation strategy provides an estimate of the local average treatment effect for ``compliers,'' defined as those players who were exposed to toxicity because they interacted with other players who were more likely to use toxic language in previous matches with other players and, consequently, were exogenously more likely to use such language in the current game. This excludes players who seek to alter their exposure to toxic language by intentionally disabling the audio chat to evade it or using toxic language to provoke reactions from other players, for instance. If the effect of exposure to toxic language were heterogeneous, this local average treatment effect might not accurately reflect the average treatment effect for the entire player population.

\subsection{Estimation}

Our model contains player-specific intercepts, formally called fixed effects, capturing the natural tendency of players to exhibit outcomes of interest. Estimation of these fixed effects is computationally expensive. Therefore, analysts frequently resort to ``down-sampling,'' which consists of sampling a computationally convenient number of observations and estimating the model with fixed effects only for those. This leads to a lower statistical accuracy. However, we do not need to compute them explicitly, especially since these fixed effects are not of primary interest to our analysis. The reason for including them in our model is to absorb time-invariant variables that affect individual players' propensity to display the outcomes of interest. This is particularly important if there is a correlation between a player's inherent tendency to display the outcomes of interest and their likelihood of being exposed to toxicity. Instead of explicitly estimating fixed effects, we can achieve the same end by demeaning the values of the dependent, independent, and instrumental variables for all players at the individual level \citep{greene_2018}. After doing so, we can estimate the coefficients $\bm{\beta}$ through the standard 2SLS estimation procedure.

We restrict our analysis to observations for which: (i) we observe at least one other player in the current match play at least another match with other players so that we can compute the value of the instruments for them, and (ii) we observe the player participate in at least two matches so that we can demean the values of the dependent, independent, and instrumental variables for them. These restrictions lead to some attrition.

\section{Results}

\begin{figure}[!p]
    \centering
    \begin{subfigure}[b]{0.9\linewidth}
        \centering
        \includegraphics[width=\linewidth]{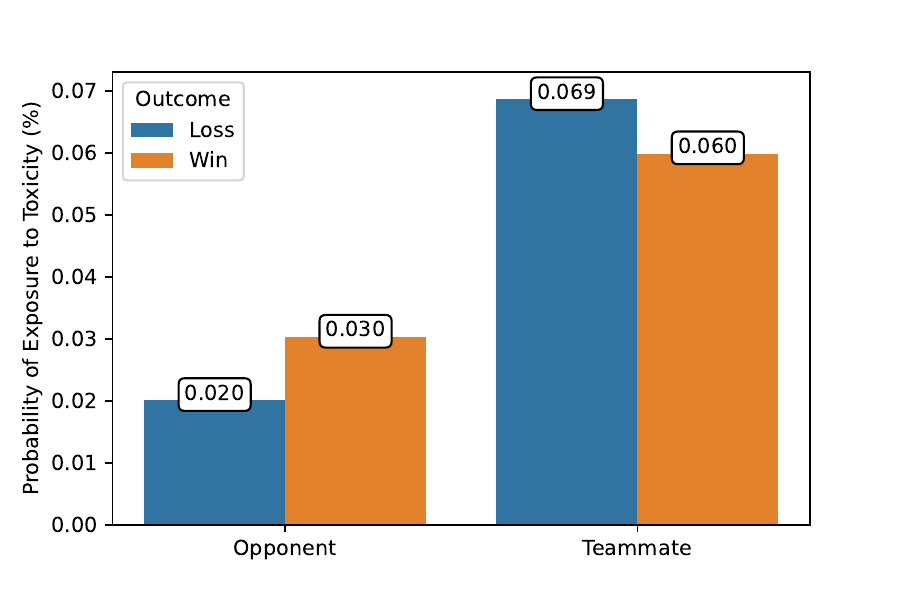}
        \caption{Opponents and Teammates}
        \label{subfig:pe_opponents_teammates}
    \end{subfigure}

    \begin{subfigure}[b]{0.9\linewidth}
        \centering
        \includegraphics[width=\linewidth]{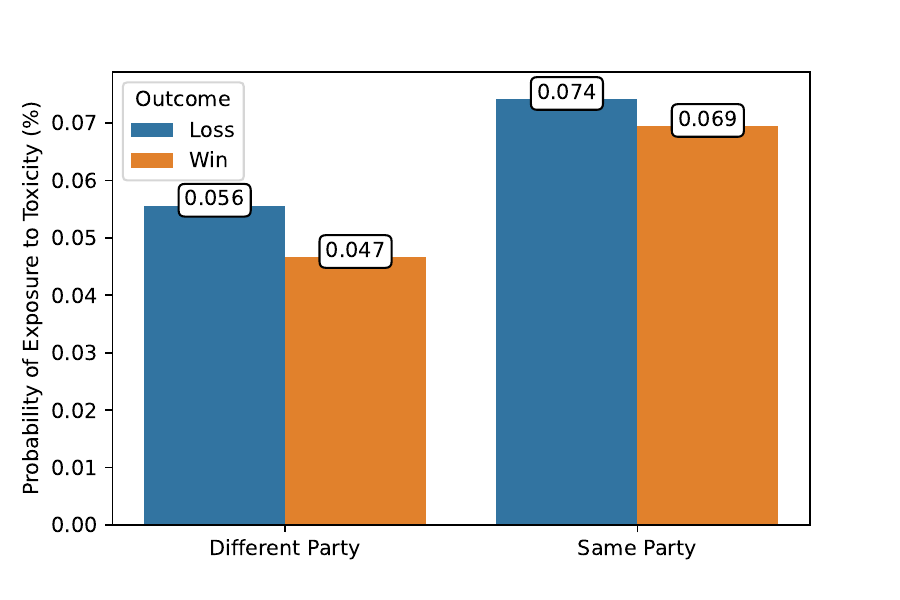}
        \caption{Teammates in a Different Party and the Same Party}
        \label{subfig:pe_teammates}
    \end{subfigure}
    \caption{Probability of Exposure to Toxic Language}
    \label{fig:exposure}
\end{figure}

\begin{figure}[!p]
    \centering
    \begin{subfigure}[b]{0.9\linewidth}
        \centering
        \includegraphics[width=\linewidth]{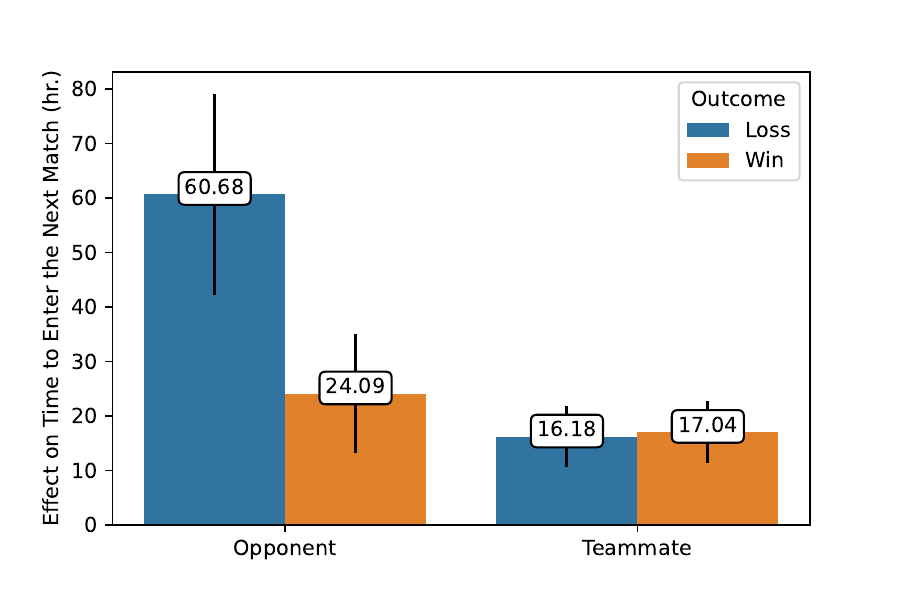}
        \caption{Time it Takes a Player to Enter the Next Match}
        \label{subfig:t_vs_o_time}
    \end{subfigure}
    
    \begin{subfigure}[b]{0.9\linewidth}
        \centering
        \includegraphics[width=\linewidth]{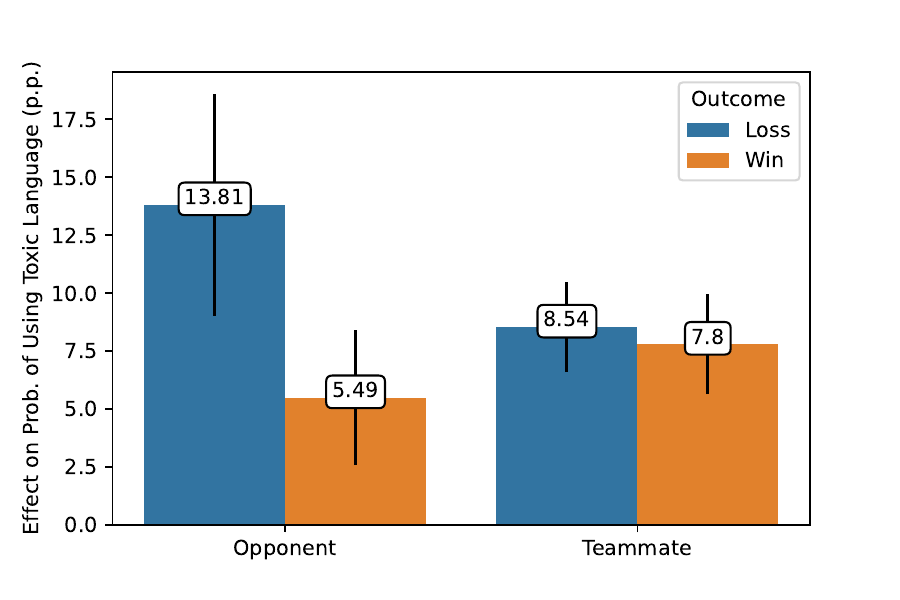}
        \caption{Probability That a Player Uses Toxic Language in the Current Match}
        \label{subfig:t_vs_o_peer}
    \end{subfigure}
    \caption{Effect of Exposure to Toxic Language from Opponents and Teammates}
    \label{fig:teammates_vs_opponents}
\end{figure}

\begin{figure}[!p]
    \centering
    \begin{subfigure}[b]{0.9\linewidth}
        \centering
        \includegraphics[width=\linewidth]{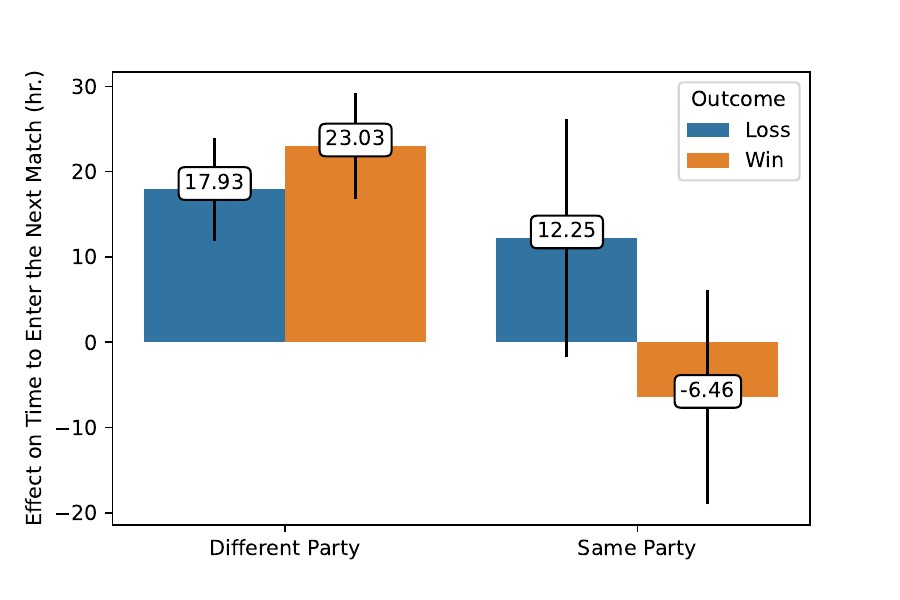}
        \caption{Time it Takes a Player to Enter the Next Match}
        \label{subfig:teammates_time}
    \end{subfigure}
    
    \begin{subfigure}[b]{0.9\linewidth}
        \centering
        \includegraphics[width=\linewidth]{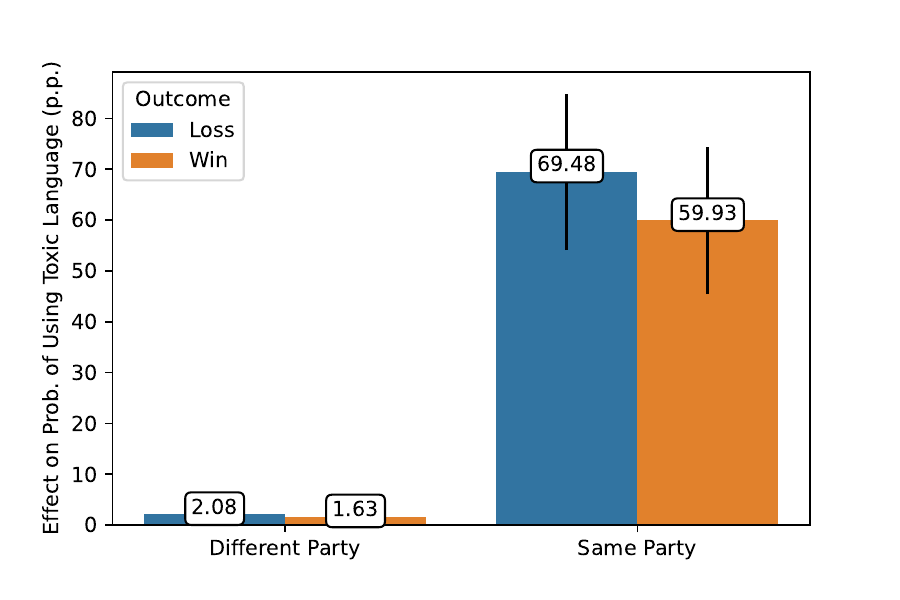}
        \caption{Probability That a Player Uses Toxic Language in the Current Match}
        \label{subfig:teammates_peer}
    \end{subfigure}
    \caption{Effect of Exposure to Toxic Language from Teammates in a Different Party and the Same Party}
    \label{fig:teammates}
\end{figure}

The results of our analysis are outlined in Figures \ref{fig:exposure} through \ref{fig:teammates}. We discuss each in turn.

\subsection{Probability of Exposure to Toxic Language}

To begin, we consider the probability that players are exposed to toxic language. Figure \ref{fig:exposure} illustrates the observed probability that players are exposed to toxic language in a match contingent on the match's outcome, defined as whether their team won or lost. In particular, Figure \ref{subfig:pe_opponents_teammates} illustrates the probability that players are exposed to toxic language from opponents and teammates, and Figure~\ref{subfig:pe_teammates} the probability that players are exposed to toxic language from teammates in a different party and the same party.

The observed probability that players are exposed to toxic language from opponents or teammates is lower than one-tenth of one percent. Though some exposure data is unavailable, thus the actual exposure to toxic language could be slightly higher, this indicates that ToxMod detects the most severe instances of toxicity.\footnote{For further details on unavailable exposure data, see the Methods section.} Players are considerably more likely---two to over three times more likely, depending on the match’s outcome---to be exposed to toxic language from teammates than from opponents. Also, players are more likely to be exposed to toxic language from opponents when their team wins and from teammates when their team loses the match. Among teammates, there is a slightly higher probability that players are exposed to toxic language from teammates in the same party. However, this difference in the probability of exposure to toxic language from teammates in a different party and the same party is much smaller than the difference in the probability of exposure to toxic language from opponents and teammates.

\subsection{Effect of Toxic Language from Opponents and Teammates}

We now turn to the effect of exposure to toxic language on player engagement and their probability of using similar language in the current match. Figure \ref{fig:teammates_vs_opponents} illustrates the estimated effect of exposure to toxic language from opponents and teammates contingent on the match's outcome. Specifically, Figure \ref{subfig:t_vs_o_time} illustrates the effect on the time it takes players to enter their next game, and Figure \ref{subfig:t_vs_o_peer} the effect on the probability that players use toxic language themselves. Each estimate represents the average marginal effect of exposure to toxic language from one player---either opponent or teammate---on the outcomes of interest. Estimates are illustrated with their 95\% confidence interval. The corresponding regression table can be found in the Appendix.

Exposure to toxic language significantly increases the average time it takes players to enter their next match. This effect ranges from 16.18 to 60.68 hours, depending on whether the toxic language originates from opponents or teammates and whether the player’s team won or lost. The effect of exposure to toxic language from opponents is larger when the player's team loses. In contrast, the effect of exposure to toxic language from opponents is greater when the player's team wins. The most pronounced effect is caused by exposure to toxic language from opponents when the player’s team loses. The effect of exposure to toxic language from opponents when the player’s team wins is much smaller. The latter is not significantly different---although slightly greater in magnitude---than the effect of exposure to toxic language from teammates regardless of the match’s outcome.

Exposure to toxic language increases the probability that a player uses similar language in the current match. The effect ranges from 5.49 to 13.81 percentage points, depending on whether the toxic language originates from opponents or teammates and whether the player’s team won or lost. This effect is substantial in magnitude, especially given that the observed incidence of toxic language is 0.083\%. Irrespective of whether it comes from opponents or teammates, the effect of exposure to toxic language is greater when the player’s team loses. The most pronounced effect is induced by exposure to toxic language from opponents when the player’s team loses. In contrast, the least pronounced effect is induced by exposure to toxic language from opponents when the player’s team wins. The latter is significantly smaller than the former. Exposure to toxicity from teammates exerts an effect of intermediate value on the probability that a player resorts to a similar language.

\subsection{Effect of Toxic Language from Different-Party and Same-Party Teammates}

Figure \ref{fig:teammates} illustrates the estimated effect of exposure to toxicity from teammates in a different party and the same party contingent on the match's outcome. Specifically, Figure \ref{subfig:teammates_time} illustrates the effect on the time it takes players to enter their next game, and Figure \ref{subfig:teammates_peer} the effect on the probability that a player resorts to toxic language themselves. Each estimate represents the average marginal effect of being exposed to toxic language from one player---either a teammate in a different party or the same party---on the outcomes of interest. Estimates are illustrated with their 95\% confidence interval. The corresponding regression table can be found in the Appendix.

Exposure to toxic language from teammates in a different party significantly increases the average time it takes a player to enter their next match. The effect is sizable, with a delay of 17.938 hours after a loss and 23.04 hours after a win. In contrast, exposure to toxic language from teammates in the same party does not significantly affect the time it takes to enter the next match, irrespective of the match’s outcome. In particular, the effect of exposure to toxic language from teammates in the same party is negative when their team wins and significantly smaller than the effect of exposure to toxicity from teammates in a different party.

Exposure to toxic language increases the probability that a player resorts to a similar language. The effect of exposure to toxic language from teammates in the same party is significant and pronounced, resulting in a 59.93 to 69.48 percentage point increase in the probability that a player resorts to toxic language depending on the match's outcome. In contrast, the effect of exposure to toxic language from teammates in a different party is much smaller, with a magnitude of 1.63 to 2.08 percentage points depending on the match's outcome. Again, when a player's team loses, assuming all other factors remain the same, toxic language tends to propagate more as compared to when the team wins.

\section{Discussion}

Our analysis yields valuable insights into the effect of exposure to toxic language on player engagement and the probability of using similar language. Our findings confirm that exposure to toxic language considerably affects player engagement, usually in a detrimental manner. Further, our findings reveal that toxic language tends to propagate, as exposure to it causes other players to use similar language. Overall, this suggests that video game publishers have a vested interest in addressing toxicity.

We demonstrate that the effect of exposure to toxic language varies depending on whether it emanates from opponents or teammates, whether it emanates from teammates in the same party or a different party, and the match’s outcome. These findings have meaningful implications regarding how resources for addressing toxicity should be allocated.

To minimize the adverse effects of toxic language on player engagement, our analysis suggests that efforts and resources dedicated to addressing toxicity should be allocated in the following order of decreasing priority:
\begin{enumerate}
    \item Toxic language from opponents when the player’s team loses.
    \item Toxic language from opponents when the player’s team wins.
    \item Toxic language from teammates in a different party when the player’s team wins.
    \item Toxic language from teammates in a different party when the player’s team loses.
    \item Toxic language from teammates in the same party when the player’s team loses.
    \item Toxic language from teammates in the same party when the player’s team wins.
\end{enumerate}

On the other hand, to minimize the proliferation of toxic language, our analysis suggests that efforts and resources devoted to addressing toxicity should be allocated in the following order of decreasing priority:
\begin{enumerate}
    \item Toxic language from teammates in the same party when the player’s team loses.
    \item Toxic language from teammates in the same party when the player’s team wins.
    \item Toxic language from opponents when the player’s team loses.
    \item Toxic language from opponents when the player’s team wins.
    \item Toxic language from teammates in a different party when the player’s team loses.
    \item Toxic language from teammates in a different party when the player’s team wins.
\end{enumerate}

To some extent, our conclusions diverge depending on whether the goal is primarily to minimize the detrimental impact of exposure to toxic language on player engagement or the propagation of toxic language. If the goal is to curtail the adverse effects of toxic language on player engagement, toxic language from opponents should be addressed as a priority. On the other hand, toxic language from teammates in the same party has the lowest effect on player engagement. However, it is precisely toxic language from teammates in the same party that results in the most propagation and, therefore, should be addressed as a priority if our goal is to curtail this propagation. In all cases, toxic language tends to have a greater effect, all else being equal, when a player's team loses, hence this is in this context that resources should be allocated in priority.

\clearpage

\section*{Acknowledgements}

The authors thank Andrea Boonyarungsrit, Grant Cahill, Min Kim, Rafal Kocielnik, Jonathan Lane, Zhuofang Li, Gary Quan, Deshawn Sambrano, Feri Soltani, Carly Taylor, and Michael Vance for their assistance and feedback.

\section*{Author Contributions}

AM and JM developed the study's methodology, and JM conducted the analyses. JM drafted the manuscript with contributions from all co-authors. RMA managed the project. All authors have accepted responsibility for the entire content of this manuscript, consented to its submission to the journal, reviewed all the results, and approved the final version of the manuscript.

\section*{Competing Interests}

Activision provided funding for this study through a sponsored research grant. AM contributed to this article while being employed by Activision. The opinions expressed by the authors do not represent the views of Activision. The other authors state that they have no competing interests.

\section*{Data Availability}

The data analyzed in this study contains proprietary information owned by Activision. Access to this data is restricted. For inquiries regarding access, please contact R. Michael Alvarez at \href{mailto:rma@hss.caltech.edu}{rma@hss.caltech.edu}.

\section*{Ethical Statements}

\subsection*{Ethical Approval}

This article does not contain any studies with human participants performed by any of the authors. This research received an exemption from Caltech's Institutional Review Board. All procedures adhered to applicable guidelines and regulations.

\subsection*{Informed Consent} 

This article does not contain any studies with human participants performed by any of the authors.

\clearpage

\printbibliography

\clearpage

\section*{Appendix}

\begin{table}[!h]
    \centering
    \caption*{\textbf{Table:} Regression Results}
    \label{tab:my_label}
    \begin{subtable}[b]{\linewidth}
        \centering
        \caption{Opponents and Teammates}
        \footnotesize
        \begin{tabular}{lcccc}
            \hline
            \hline
            \\[-1.5ex]
            & \textbf{Time to Enter the Next Match} & \textbf{Prob. of Using Toxic Language} \\
            \\[-1.5ex]
            \hline
            \\[-1.5ex]
            Opponents & 60.683$^{***}$ & 0.1382$^{***}$ \\
            & (9.4202) & (0.0244) \\
            \\[-1.5ex]
            Teammates & 16.182$^{***}$ & 0.0854$^{***}$ \\
            & (2.8458) & (0.0099) \\
            \\[-1.5ex]
            Opponents $\times$ Win & $-$36.596$^{***}$ & $-$0.0832$^{***}$ \\
            & (10.781) & (0.0281) \\
            \\[-1.5ex]
            Teammates $\times$ Win & 0.8545 & $-$0.0074 \\
            & (3.8833) & (0.0152) \\
            \\[-1.5ex]
            Win & $-$0.3794$^{***}$ & $-$0.0001$^{***}$ \\
            & (0.0053) & (0.0000) \\
            \\[-1.5ex]
            \hline
            \\[-1.5ex]
            $F$ Statistic & 8,958.2 & 504.1 \\
            \\[-1.5ex]
            \hline
            \hline
            \\[-2ex]
            \textit{Note:}  & \multicolumn{2}{r}{$^{*}$p$<$0.1; $^{**}$p$<$0.05; $^{***}$p$<$0.01}
        \end{tabular}
    \end{subtable}

    \bigskip

    \begin{subtable}[b]{\linewidth}
        \centering
        \caption{Teammates in a Different Party and the Same Party}
        \footnotesize
        \begin{tabular}{lcccc}
            \hline
            \hline
            \\[-1.5ex]
            & \textbf{Time to Enter the Next Match} & \textbf{Prob. of Using Toxic Language} \\
            \\[-1.5ex]
            \hline
            \\[-1.5ex]
            Different Party & 17.936$^{***}$ & 0.0208$^{***}$ \\
            & (3.0740) & (0.0078) \\
            \\[-1.5ex]
            Same Party & 12.255$^{*}$ & 0.6949$^{***}$ \\
            & (7.1299) & (0.0785) \\
            \\[-1.5ex]
            Different Party $\times$ Win & 5.0996 & $-$0.0045 \\
            & (4.3649) & (0.0111) \\
            \\[-1.5ex]
            Same Party $\times$ Win & $-$18.705$^{**}$ & $-$0.0956 \\
            & (7.5382) & (0.0910) \\
            \\[-1.5ex]
            Win & $-$0.3811$^{***}$ & $-$0.0001$^{***}$ \\
            & (0.0048) & (0.0000) \\
            \\[-1.5ex]
            Player Belong to Party & $-$0.4193$^{***}$ & 0.0013$^{***}$ \\
            & (0.0082) & (0.0000) \\
            \\[-1.5ex]
            \hline
            \\[-1.5ex]
            $F$ Statistic & 12,551.2 & 8,454.4 \\
            \\[-1.5ex]
            \hline
            \hline
            \\[-2ex]
            \textit{Note:}  & \multicolumn{2}{r}{$^{*}$p$<$0.1; $^{**}$p$<$0.05; $^{***}$p$<$0.01}
        \end{tabular}
    \end{subtable}
\end{table}

\end{document}